\documentclass[pra,onecolumn,floatfix,superscriptaddress,longbibliography,notitlepage, nofootinbib]{revtex4-1}
\pdfoutput=1

\usepackage[utf8]{inputenc}
\usepackage{braket}
\usepackage{amsmath}
\DeclareMathOperator{\Tr}{Tr}
\usepackage{dcolumn}   
\usepackage{bm}        
\usepackage{amssymb}
\usepackage{mathtools}
\usepackage{tikz}
\usepackage{qcircuit}
\usepackage{appendix}
\usepackage{hyperref}

\usepackage{subcaption}
\usepackage{amsmath,amsfonts,amssymb}
\usepackage{mathtools}
\usepackage{thmtools,thm-restate}
\usepackage{algorithm}

\usepackage{algpseudocode}

\newtheorem{theorem}{Theorem}

\newtheorem{method}{Algorithm}

\usetikzlibrary{arrows.meta}

\def\BibTeX{{\rm B\kern-.05em{\sc i\kern-.025em b}\kern-.08em
    T\kern-.1667em\lower.7ex\hbox{E}\kern-.125emX}}

\begin{document}
\title{A Framework For Estimating Amplitudes of Quantum State With Single-Qubit Measurement }
\author{Nhat A. Nghiem}
\affiliation{Department of Physics and Astronomy, State University of New York at Stony Brook, Stony Brook, NY 11794-3800, USA}
\affiliation{C. N. Yang Institute for Theoretical Physics, State University of New York at Stony Brook, Stony Brook, NY 11794-3840, USA}

\begin{abstract}
    We propose and analyze a simple framework for estimating the amplitudes of a given $n$-qubit quantum state $\ket{\psi} = \sum_{i=0}^{2^n-1} a_i \ket{i}$ in computational basis, utilizing a single-qubit measurement only. Previously, it was a common procedure that one could measure all qubits in order to collect measurement outcomes, from which one can estimate amplitudes of given quantum state. Here, we show that if restricting to single-qubit measurement, and one can perform measurement on arbitrary basis, then the measurement outcomes can be used to assist the finding of amplitudes in the usual computational, or Z basis. More concretely, such outcomes are capable of constructing a system of nonlinear algebraic equations, and by classically solving them, we obtain $\Tilde{a}_i$, which is the approximation to the corresponding amplitudes $a_i$, including both real and imaginary component. We then discuss our framework from a broader perspective. First, we show that estimating all (norms of) amplitudes to additive accuracy $\delta$, i.e., $| |\Tilde{a}_i - |a_i| | \leq \delta$ for all $i$, $\mathcal{O}(4^n/\delta^4)$ single-qubit measurements is sufficient. Second, we show that to achieve total variation $\sum_{i=0}^{2^n-1} | |\Tilde{a}_i|^2 - |a_i|^2| \leq \delta $, $\mathcal{O}(6^n/\delta^4)$ a single bit measurement is required. Finally, in order to achieve an average $L_1$ norm error $ \sum_{i=0}^{2^n-1} | |\Tilde{a}_i| - |a_i| |/2^n  \leq \delta$, a single bit measurement $\mathcal{O}(2^n/ \delta^4)$ is needed. 
\end{abstract}
\maketitle

\section{Introduction}
Quantum computation has been undergoing rapid development since some early proposals \cite{feynman2018simulating, deutsch1985quantum, deutsch1992rapid, lloyd1996universal, shor1999polynomial, grover1996fast}. By harnessing intrinsic properties of quantum physics, such as entanglement and superposition, a quantum computer can store, handle, and process information in a very different way compared to classical devices. The principle of quantum computation can be explained simply by the linear algebraic language. Typically, in an usual model of quantum computation, one begins with a state, which is a vector in a Hilbert space, then applies a sequence of gates, which are unitary matrices, followed by measurement operation at the end. The measurement operation features another unique property of quantum physics, where the system, represented by some state vector, is projected, in a probabilistic manner, into some other state depending on the kind of measurement operation being used. As a quantum state is represented by a vector in a Hilbert space, one can choose a basis, and in theory, any state admits decomposition in such a basis. However, given a basis, it is challenging to characterize, i.e. extracting the coefficients of any given unknown state, as the dimension, or parameters, grows exponentially with respect to the number of constituents, e.g., qubits. Furthermore, these parameters cannot be directly accessed but can only be revealed by performing measurement. The results of the measurement are then used to estimate the desired parameters. As mentioned, the measurement operation projects the given state into another one, an occurrence referred to as wave-function collapse. Hence, to characterize a given state completely, we need to perform many measurements on multiple copies of the given state. 

Several protocols have been proposed to address the challenge and related problem mentioned above. For example, references \cite{huang2020predicting} and \cite{huang2021efficient} introduce learning-based methods to estimate linear properties, such as Pauli observables, of a given density state $\rho$. More relevant to our main objective in this work is quantum state tomography \cite{yu2020sample,bagan2004collective,keyl2006quantum,guctua2008optimal,haah2016sample,o2016efficient, guctua2020fast,flammia2012quantum,kueng2017low}, where measurement, including single-qubit measurements and multi-qubit measurements on copies of a given state $\rho$, or could possibly be $\rho^{\otimes k}$ for some $k$, is carried out to obtain a classical description of given $\rho$. 

Here, we observe that single-qubit measurement essentially contains more insights that could be more informative and beneficial for probing a given quantum system. A common measurement that previous work \cite{yu2020sample, bagan2004collective} used is the Pauli measurement, corresponding to a certain basis of measurement. One can ask, given the vastness of the Hilbert space where a quantum state vector resides, that in principle there are countless basis of measurement, so what if we can utilize another basis for the purpose of characterizing a given state ? Such insight is indeed simple and informative enough to allow us to achieve the goal. The probabilistic mechanism behind measurement obeys Born's rule, and given a basis of measurement, one can explicitly write out the probability of obtaining certain outcomes (corresponding with a certain state after measurement) as a function of given state's amplitudes. The process of measuring (multiple copies of) a given state then amounts to learning discrete probability distributions, and one can statistically estimate the probabilities corresponding to each outcome. These estimations are then used to reveal the given state.

More concretely, we consider an $n$-qubit system $\ket{\psi} = \sum_{i=0}^{2^n-1} a_i \ket{i}$ where $\ket{i}$ is a computational basis state (or commonly known as Pauli $Z$-basis), and we are interested in knowing all the so-called amplitudes $\{ a_i\}_{i=0}^{2^n-1}$, which are generally complex numbers. It is known that if we measure all qubits on the basis of $Z$, then the probability of obtaining $\ket{i}$ after measurement is $\Tr( \ket{\psi}\bra{\psi} \cdot \ket{i}\bra{i}) = |a_i|^2$. Then using Hoeffding inequality, we can measure $\ket{\psi}$ $\mathcal{O}( 1/\epsilon^2    )$ times (using different copies) to estimate $|a_i|^2$ to accuracy $\epsilon$. The whole process is then repeated to estimate all the amplitude squares to the desired accuracy. Our protocol introduced in this work aims to rectify this task and closely related task (to be elaborated later), using single-qubit measurement only. As mentioned, we utilize the vastness of Hilbert space and that different bases of measurement are possible. Once a measurement (with chosen basis) is placed on a qubit, then the probability of obtaining a certain outcome can be derived explicitly. As we show explicitly later, this probability with each outcome could be expressed as a function of all amplitudes $\{a_i\}_{i=0}^{2^n-1}$. In addition, these probabilities could be estimated statistically via measurement, and all such probabilities then form a system of nonlinear algebraic equations, by which solving them, the desired amplitudes could be estimated. As mentioned, our method employs only single-qubit measurement. Thus, it shares similarity with previous work \cite{yu2020sample}. The difference is in the basis of measurement, as we go beyond the Pauli measurement.  

In the following section \ref{sec: examples} we begin with two examples for 2-qubit and 3-qubit case to illustrate the insight behind our method. From these examples, generalization to $n$-qubit case is discussed subsequently \ref{sec: generalframework}. In Section \ref{sec: additiveerror}, \ref{sec: totalvariance}, \ref{sec: estimatingl1norm}, we discuss our frameworks from a broader perspective, where we explicitly show the complexity, or the scaling of measurements, with respect to different expectation of error. We conclude in section \ref{sec: conclusion} and discuss the potential future direction stemming from this work. 

\section{Main Framework}
\label{sec: amplitudeestimation}
We first provide two illustrating examples in section \ref{sec: examples}, followed by a general framework provided in section \ref{sec: generalframework}. 
\subsection{Examples}
\label{sec: examples}
We begin by discussing two examples: $\ket{\psi}_1 \in \mathbb{C}^4$ being the quantum state of 2-qubits and $\ket{\psi}_2 \in \mathbb{C}^8$ being the quantum state of 3-qubits. 

Let $\ket{\psi}_1 = a_{00} \ket{00} + a_{01} \ket{01} + a_{10} \ket{10} + a_{11} \ket{11}$ be the decomposition of $\ket{\psi}_1$ in the computational basis state. If we measure two qubits, then it is fundamental, that is, Born's rule, that the probability of measuring string $(i,j)$ ($i,j=0,1$) is $p_{ij} = \Tr( \ket{ij}\bra{ij} \cdot \ket{\psi}_1 \bra{\psi}_1) = |a_{ij}|^2$. If we want to estimate all of this probability $p_{ij}$, then we need to repeat the measurement multiple times and use the results to statistically estimate all probabilities $p_{ij}$. 

To see how the outcomes can be used for estimating all probabilities $p_{ij}$ from measuring $\ket{\psi}_1$ in computational basis, we can first consider the estimation of $p_{00}$. Then the measurement outcomes form a Bernoulli distribution for $00$, as we simply treat all the outcomes $\{ 01,10,11\}$ as a discrete variable with corresponding probability $1-p_{00}$. It is known (via Chernoff bound) that the samples required to estimate $p_{00}$ to accurary $\epsilon$ with probability at least $1-\delta$ is $\mathcal{O}(\ln(1/\delta) /\epsilon^2)$. The same strategy can be applied to estimate $p_{01},p_{10}, p_{11}$ by treating each of them and the remaining probabilities as Bernoulli distribution. In the end, the whole procedure takes $\mathcal{O}(4 \ln(1/\delta)/\epsilon^2)$ repetitions to estimate all probabilities to additive accuracy $\epsilon$, with high probability.  

The same result is applied for 3-qubits case. Let $\ket{\psi}_2 = a_{000} \ket{000} + a_{001} \ket{001} + a_{010} \ket{010} + a_{011} \ket{011} + a_{100} \ket{100} + a_{101}\ket{101} + a_{110} \ket{110} + a_{111} \ket{111}$. Then repeating the measurements in computational basis state allows us to estimate each probability $p_{ijk} = |a_{ijk}|^2$ (for $i,j,k=0,1$) to error at most $\epsilon$ with success probability at least $1-\delta$, with $\mathcal{O}(8 \ln(1/\delta)/\epsilon^2)$ repetitions according to the above lemma. It is straightforward to see that even in $n$-qubits case, the procedure is the same for estimating all amplitudes, hence the sample complexity is $\mathcal{O}(2^n \ln(1/\delta)/\epsilon^2)$ to achieve the desired precision. 

Now for $\ket{\psi}_1$, instead of measuring all qubits in computational basis state (with four possible outcomes $00,01,10,11$), we measure the first qubit in computational basis state. Then there will be two possible outcomes, which are 0 and 1 (or more like $\ket{0}$ or $ \ket{1}$ in the first qubit), with corresponding probability $p_0$ and $p_1$:
\begin{align}
    p_{0} = \Tr(  \ket{0}\bra{0} \otimes \mathbb{I}_{2} \cdot \ket{\psi}_1\bra{\psi}_1 ) \\
    p_{1} = \Tr(  \ket{1}\bra{1} \otimes \mathbb{I}_{2} \cdot \ket{\psi}_1\bra{\psi}_1 ) 
\end{align}
where $\mathbb{I}_2$ refers to the identity matrix on the second qubit system. To avoid the confusion, as subsequently we also have $p_0$ and $p_1$ but those probabilities are for the measurement's outcome on the second qubit, we use the superscript to denote the qubit explicitly: $p_0^{1 (or 2)}$ and $p_1^{1 (or 2)}$ refers to the probability of obtaining $0$ and $1$ on the first, or second qubit respectively.  We remark the following crucial information, which is essentially a marginal probability property:
\begin{align}
    p_{00} + p_{01} =  p_0^1\\
    p_{10} + p_{11} = p_1^1 
\end{align}
where $p_{ij} = |a_{ij}|^2$ is the usual squared amplitude of state $\ket{\psi}_1$. Likewise, if we measure the second qubit, there would also be two possible outcomes 0 and 1, with probability $p_0^2$ and $p_1^2$. Then we also have that, for second qubit: 
\begin{align}
    p_{00} + p_{10} = p_0^2 \\
    p_{01} + p_{11} = p_1^2
\end{align}
It is straightforward to see that the above equation form a linear system:
\begin{align}
\label{eqn: 2qubitlinear}
    \begin{pmatrix}
        1 & 1 & 0 & 0 \\
        0 & 0 & 1 & 1 \\
        1 & 0 & 1 & 0 \\
        0 & 1 & 0 & 1 
    \end{pmatrix} \begin{pmatrix}
        p_{00} \\
        p_{01} \\
        p_{10}\\
        p_{11}
    \end{pmatrix} = \begin{pmatrix}
        p_0^1 \\
        p_1^1 \\
        p_0^2 \\
        p_1^2 
    \end{pmatrix}
\end{align}
As we have discussed, $\mathcal{O}( \ln(1/\delta) /\epsilon^2)$ repetition is required to estimate $p_0^1,p_0^2$ and $p_1^1, p_1^2$ to the precision $\epsilon$. This means that the right-hand side of the above linear system is erroneous. Therefore, if we wish to solve the linear system above to obtain the probability $p_{00}, p_{01}, p_{10}, p_{11}$, then the solution appears to be erroneous, which we will discuss in more detail later. 

Now we examine the same procedure as above for 3-qubits cases. We measure each qubit of the state $\ket{\psi}_2$ $\mathcal{O}(1/\epsilon^2)$ times, and obtain the outcomes $p_0^1, p_1^1, p_0^2, p_1^2, p_0^3, p_1^3$. Recall that 
\begin{align}
    \ket{\psi}_2 &= a_{000} \ket{000} + a_{001} \ket{001} + a_{010} \ket{010} + a_{011} \ket{011} + a_{100} \ket{100} + a_{101}\ket{101} + a_{110} \ket{110} + a_{111} \ket{111} 
\end{align} 
where each amplitude squared $|a_{ijk}|^2 = p_{ijk}$ is the probability of obtaining corresponding outcome $(i,j,k)$. Then due to marginal probability property, we have that:
\begin{align}
    p_{000} + p_{001} + p_{010} + p_{011} = p_0^1 \\
    p_{100} + p_{101} + p_{110} + p_{111} = p_1^1 \\
    p_{000} + p_{001} + p_{100} + p_{101} = p_0^2 \\
    p_{010} + p_{011} + p_{110} + p_{111} = p_1^2 \\
    p_{000} + p_{010} + p_{100} + p_{110} = p_0^3 \\
    p_{001} + p_{011} + p_{101} + p_{111} = p_1^3 
    \label{eqn: A}
\end{align}
The above equations can be written as a linear system:
\begin{align}
    \begin{pmatrix}
        1 & 1 & 1 & 1 & 0 & 0 & 0 & 0 \\
        0 & 0 & 0 & 0 & 1 & 1 & 1 & 1 \\
        1 & 1 & 0 & 0 & 1 & 1 & 0 & 0 \\
        0 & 0 & 1 & 1 & 0 & 0 & 1 & 1 \\
        1 & 0 & 1 & 0 & 1 & 0 & 1 & 0 \\
        0 & 1 & 0 & 1 & 0 & 1 & 0 & 1 
    \end{pmatrix} \begin{pmatrix}
        p_{000} \\
        p_{001} \\
        p_{010} \\
        p_{011} \\
        p_{100} \\
        p_{101} \\
        p_{110} \\
        p_{111}
    \end{pmatrix} = \begin{pmatrix}
        p_0^1 \\
        p_1^1 \\
        p_0^2 \\
        p_1^2 \\
        p_0^3 \\
        p_1^3
    \end{pmatrix}
    \label{eqn: 3qubitsystem}
\end{align}
The above system is rectangular, rather than square system comparing to Eqn. \ref{eqn: 2qubitlinear}, which makes it a bit trickier to solve, as there are more variables than the number of equations. In the context of linear algebra, it is referred as underdetermined systems, and unique solution might not exists, rather, there can be infinitely many solutions. To make it square, we need to have more equations. However, the used procedure employs a single qubit measurement only, and we already measure all individual qubit. In order to collect more information as to obtain more equations, we need to have more measurement outcomes.

There are two possible ways to solve the aforementioned challenge. The first one is, we can perform joint measurement on $2$ qubits. Suppose we perform measurement on the first two qubits of $\ket{\psi}_2$, then there are four possible outcomes: 00,01,10,11. Let $p_{ij}^{1,2}$ with $i,j=0,1$ denote the probability of measuring $(i,j)$ on the qubit 1 and 2 of $\ket{\psi}_2$. Then because of the decomposition of $\ket{\psi}_2 = a_{000} \ket{000} + a_{001} \ket{001} + a_{010} \ket{010} + a_{011} \ket{011} + a_{100} \ket{100} + a_{101}\ket{101} + a_{110} \ket{110} + a_{111} \ket{111}$, and $|a_{ijk}|^2 = p_{ijk}$ is the corresponding probability of measuring outcome $(i,j,k)$. Using marginal probability property as in previous 2-qubit case, we have that:
\begin{align}
    p_{000} + p_{001} = p_{00}^{1,2} \\
    p_{010} + p_{011} = p_{01}^{1,2} 
\end{align}
Note that there are two other outcomes $(1,0)$ and $(1,1)$, however, we don't need to use them, as the above equations, once combined with equations Eqn.\ref{eqn: A}, yields: 
\begin{align}
    p_{000} + p_{001} + p_{010} + p_{011} = p_0^1 \\
    p_{100} + p_{101} + p_{110} + p_{111} = p_1^1 \\
    p_{000} + p_{001} + p_{100} + p_{101} = p_0^2 \\
    p_{010} + p_{011} + p_{110} + p_{111} = p_1^2 \\
    p_{000} + p_{010} + p_{100} + p_{110} = p_0^3 \\
    p_{001} + p_{011} + p_{101} + p_{111} = p_1^3 \\
    p_{000} + p_{001} = p_{00}^{1,2} \\
    p_{010} + p_{011} = p_{01}^{1,2} 
\end{align}
which is a square linear system:
\begin{align}
    \begin{pmatrix}
        1 & 1 & 1 & 1 & 0 & 0 & 0 & 0 \\
        0 & 0 & 0 & 0 & 1 & 1 & 1 & 1 \\
        1 & 1 & 0 & 0 & 1 & 1 & 0 & 0 \\
        0 & 0 & 1 & 1 & 0 & 0 & 1 & 1 \\
        1 & 0 & 1 & 0 & 1 & 0 & 1 & 0 \\
        0 & 1 & 0 & 1 & 0 & 1 & 0 & 1 \\
        1 & 1 & 0 & 0 & 0 & 0 & 0 & 0 \\
        0 & 0 & 1 & 1 & 0 & 0 & 0 & 0 
    \end{pmatrix} \begin{pmatrix}
        p_{000} \\
        p_{001} \\
        p_{010} \\
        p_{011} \\
        p_{100} \\
        p_{101} \\
        p_{110} \\
        p_{111}
    \end{pmatrix} = \begin{pmatrix}
        p_0^1 \\
        p_1^1 \\
        p_0^2 \\
        p_1^2 \\
        p_0^3 \\
        p_1^3 \\
        p_{00}^{1,2} \\
        p_{01}^{1,2} 
    \end{pmatrix}
\end{align}
In principle, it is sufficient from the above square system to solve for desired probability from the two qubits measurement outcomes $(0,0)$ and $(0,1)$, in addition to single qubit measurements. \\

While the above described joint qubit measurement seems efficient enough to do, however, the whole procedure gives us a linear system with probabilities, or amplitude square, as variables. We remark that, all amplitudes are complex number, so for this 3-qubit case, we have $2^3$ amplitudes, corresponding with $2^{3+1}$ complex variables. Therefore, we can only find the amplitude square, meanwhile, amplitude could be complex and the above linear system isn't giving us enough information to find the amplitude, including its real and complex part. Performing more joint 2-qubit measurements, in principle, can give us more equations. However, to our surprise, solution based on only single qubit measurement exists. Remind that, in order to obtain the equation Eqn. \ref{eqn: 3qubitsystem}, we performed single qubit measurement on $Z$ basis. In fact, we can also perform measurement on arbitrary basis, and it should be able to give us more information. To proceed, let $\theta$ denotes the angle for another basis measurement, and $\ket{\theta}_{0/1}$ be the eigenstate of corresponding measurement operator. Then the probability of obtaining $\ket{\theta}_0$ or $\ket{\theta}_1$ is:
\begin{align}
    p(\theta)_{0/1} = \Tr ( \ket{\theta}_{0/1}\bra{\theta}_{0/1} \otimes \mathbb{I}_{n-1} \cdot \ket{\psi}_2 \bra{\psi}_2   )
\end{align}
where $\mathbb{I}_{4}$ refers to identity matrix in the remaining $2$ qubits system. Let 
\begin{align}
    \ket{\theta}_0 = \cos(\theta) \ket{0} + \sin(\theta) \ket{1} \\
    \ket{\theta}_1 = \sin(\theta) \ket{0} - \cos(\theta) \ket{1}
\end{align}
be the decomposition of $\ket{\theta}_{0/1}$ in regular computational basis. From the above equation, it is easy to rewrite:
\begin{align}
    \ket{0} = ( \cos(\theta) \ket{\theta}_0 + \sin(\theta) \ket{\theta}_1 ) \\
    \ket{1} = ( \sin(\theta) \ket{\theta}_0 - \cos(\theta) \ket{\theta}_1) 
\end{align}
Recall that 
\begin{align}
    \ket{\psi}_2 &= a_{000} \ket{000} + a_{001} \ket{001} + a_{010} \ket{010} + a_{011} \ket{011} + a_{100} \ket{100} + a_{101}\ket{101} + a_{110} \ket{110} + a_{111} \ket{111} \\
    &= \ket{0} \big( a_{000} \ket{00} + a_{001} \ket{01} + a_{010} \ket{10} + a_{011} \ket{11}  \big) + \ket{1} \big( a_{100} \ket{00} + a_{101}\ket{01} + a_{110} \ket{10} + a_{111} \ket{11} \big)  \\
    &= ( \cos(\theta) \ket{\theta}_0 + \sin(\theta) \ket{\theta}_1 )\big( a_{000} \ket{00} + a_{001} \ket{01} + a_{010} \ket{10} + a_{11} \ket{11}  \big) + \\ &   ( \sin(\theta) \ket{\theta}_0 - \cos(\theta) \ket{\theta}_1)  \big( a_{100} \ket{00} + a_{101}\ket{01} + a_{110} \ket{10} + a_{111} \ket{11} \big) 
\end{align}
Let $\ket{\phi}_1 = a_{000} \ket{00} + a_{001} \ket{01} + a_{010} \ket{10} + a_{011} \ket{11} $ and $\ket{\phi}_2 = a_{100} \ket{00} + a_{101}\ket{01} + a_{110} \ket{10} + a_{111} \ket{11} $. Then we have:
\begin{align}
    \ket{\psi}_2 &=  \ket{\theta}_0 \big( \cos(\theta)  \ket{\phi}_1 + \sin(\theta) \ket{\phi}_2  \big) + \ket{\theta}_1  \big( \cos(\theta) \ket{\phi}_1 - \sin(\theta) \ket{\phi}_2  \big)
\end{align}
If we measure the first qubit, then the probability of measuring $\ket{\theta}_0, \ket{\theta}_1$ is:
\begin{align}
    p(\theta)_0 = | \cos(\theta) \ket{\phi}_1 + \sin(\theta) \ket{\phi}_2  |^2 \\
    p(\theta)_1 = | \cos(\theta) \ket{\phi}_1 - \sin(\theta) \ket{\phi}_2  |^2
\end{align}
We have that:
\begin{align}
  \cos(\theta) \ket{\phi}_1 + \sin(\theta) \ket{\phi}_2 = & ( \cos(\theta) a_{000} + \sin(\theta) a_{100})\ket{00} + \\& ( \cos(\theta) a_{001} +\sin(\theta) a_{101} )\ket{01} + \\& (\cos(\theta) a_{010} + \sin(\theta) a_{110}  ) \ket{10} + \\& ( \cos(\theta) a_{011} + \sin(\theta) a_{111} ) \ket{11}
\end{align}
Then:
\begin{align}
    | \cos(\theta) \ket{\phi}_1 +  \sin(\theta)  \ket{\phi}_2  |^2 &= | \cos(\theta)a_{000} +  \sin(\theta)  a_{100}|^2 + \\& | \cos(\theta)a_{001} +  \sin(\theta)  a_{101}  |^2 + \\& \cos(\theta) a_{010} +  \sin(\theta)  a_{110}  |^2 + \\&  |\cos(\theta) a_{011} + \sin(\theta)  a_{111}|^2 \\
\end{align}
Likewise:
\begin{align}
    | \cos(\theta) \ket{\phi}_1 - \sin(\theta)  \ket{\phi}_2  |^2 &= | \cos(\theta)a_{000} - \sin(\theta)  a_{100}|^2 + \\& | \cos(\theta) a_{001} - \sin(\theta)  a_{101}  |^2 + \\& |\cos(\theta) a_{010} - \sin(\theta)  a_{110}  |^2 + \\&  |\cos(\theta) a_{011} - \sin(\theta) 
 a_{111}|^2 \\
\end{align}
Recall that from equation Eqn. \ref{eqn: A}:
\begin{align}
    p_{000} + p_{001} + p_{010} + p_{011} = p_0^1 \\
    p_{100} + p_{101} + p_{110} + p_{111} = p_1^1 \\
    p_{000} + p_{001} + p_{100} + p_{101} = p_0^2 \\
    p_{010} + p_{011} + p_{110} + p_{111} = p_1^2 \\
    p_{000} + p_{010} + p_{100} + p_{110} = p_0^3 \\
    p_{001} + p_{011} + p_{101} + p_{111} = p_1^3 
\end{align}
if we replace $p_{ijk} = |a_{ijk}|^2$ for $i,j,k = 0,1$, we equivalently have:
\begin{align}
    |a_{000}|^2 + |a_{001}|^2 + |a_{010} |^2 + |a_{011} |^2 = p_0^1 \\
    |a_{100}|^2 + | a_{101} |^2 + |a_{110}|^2 + |a_{111}|^2 = p_1^1 \\
    |a_{000}|^2 + |a_{001}|^2 + |a_{100}|^2 + |a_{101}|^2 = p_0^2 \\
    |a_{010} |^2 + |a_{011}|^2 + |a_{110}|^2 + |a_{111}|^2 = p_1^2 \\
    |a_{000}|^2 + |a_{010}|^2 + |a_{100}|^2 + |a_{110}|^2 = p_0^3 \\
    |a_{001}|^2 + |a_{011}|^2 + |a_{101}|^2 + |a_{111}|^2 = p_1^3 
\end{align}
From the above measurement on $\theta$-angle basis, we have:
\begin{align}
    & | \cos(\theta)a_{000} + \sin(\theta)  a_{100}|^2 + | \cos(\theta)a_{001} + \sin(\theta) a_{101}  |^2 + \\& |\cos(\theta) a_{010} + \sin(\theta)  a_{110}  |^2 +  |\cos(\theta) a_{011} + \sin(\theta)  a_{111}|^2  = p(\theta)_0 \\
    & | \cos(\theta)a_{000} - \sin(\theta)  a_{100}|^2 + | \cos(\theta)a_{001} - \sin(\theta)  a_{101}  |^2 + \\& |\cos(\theta) a_{010} -\sin(\theta)  a_{110}  |^2 +  |\cos(\theta) a_{011} - \sin(\theta)  a_{111}|^2  = p(\theta)_1 \\
\end{align}
Putting everything together, we have the following set of equations:
\begin{align}
\begin{cases}
    & |a_{000}|^2 + |a_{001}|^2 + |a_{010} |^2 + |a_{011} |^2 = p_0^1 \\
    &|a_{100}|^2 + | a_{101} |^2 + |a_{110}|^2 + |a_{111}|^2 = p_1^1 \\
    &|a_{000}|^2 + |a_{001}|^2 + |a_{100}|^2 + |a_{101}|^2 = p_0^2 \\
    &|a_{010} |^2 + |a_{011}|^2 + |a_{110}|^2 + |a_{111}|^2 = p_1^2 \\
    &|a_{000}|^2 + |a_{010}|^2 + |a_{100}|^2 + |a_{110}|^2 = p_0^3 \\
    &|a_{001}|^2 + |a_{011}|^2 + |a_{101}|^2 + |a_{111}|^2 = p_1^3 \\ 
    & | \cos(\theta)a_{000} + \sin(\theta) a_{100}|^2 + | \cos(\theta)a_{001} + \sin(\theta) a_{101}  |^2 + \\& |\cos(\theta) a_{010} + \sin(\theta) a_{110}  |^2 +  |\cos(\theta) a_{011} + \sin(\theta) a_{111}|^2  = p(\theta)_0 \\
    & | \cos(\theta)a_{000} - \sin(\theta) a_{100}|^2 + | \cos(\theta)a_{001} - \sin(\theta) a_{101}  |^2 + \\& |\cos(\theta) a_{010} - \sin(\theta) a_{110}  |^2 +  |\cos(\theta) a_{011} - \sin(\theta) a_{111}|^2  = p(\theta)_1 \\
\end{cases}
\label{eqn: nonlinearequation}
\end{align}
We remind that all amplitudes $a_{ijk}$ are complex numbers, i.e., $a_{ijk} = \Re(a_{ijk}) + i \Im (a_{ijk})$. At this point, we still don't have enough information to solve because as mentioned, all amplitudes are complex number, so this 3-qubit case, we have $2^3$ amplitudes, corresponding with $2^{3+1}$ complex variables. The above equations contain 8 equations, so we need 8 more. In fact, there are many ways to collect more equations. Previously we have showed that performing joint 2-qubit measurements can also work, however, if we restrict ourselves to single qubit, then measuring a single qubit can also work well. If we measure the same first qubit in another basis, for example, different angle $\theta'$, then we will have 2 more equations, corresponding with two possible outcomes. By choosing more angles, or we can opt to measure another qubit, then we can obtain more outcomes, thus forming further equations. 

The above examples on two-qubits and three-qubits cases have illustrated the key idea behind our proposal, as instead of measuring all qubits, we only need to measure some subsets of qubits, and even a single qubit measurement is sufficient. The outcomes are then being processed classically, e.g., solving a nonlinear system, to obtain the desired amplitudes. Now we proceed to provide a formal description for our method, generalizing the above examples into $n$-qubits state.

\subsection{General Framework}
\label{sec: generalframework}
From the above examples, it is straightforward to generalize the procedure to the $n$-qubit state. As there are $2^{n}$ amplitudes and $2^{n+1}$ variables, therefore we need $2^{n+1}$ equations. Each qubit measurement yields two possible outcomes, corresponding to the equations $2$ (see the back equation \ref{eqn: 2qubitlinear}, \ref{eqn: A}), then measuring each qubit (among $n$ qubits) on a given basis yields $2n$ equations. If, as a first step, we choose to measure each individual qubit in Z basis, then we have $2n$ equations. Then, we continue to measure only the first qubit, then if in total $M$ is a different basis to measure, then we require $2M \geq 2^{n+1} -2n$. Given that $M$ is an integer, one can opt for $M = \lceil 2^{n}-n \rceil$, which is the total of different bases, or different angles, which one needs to choose and measure. 
\begin{method}
\label{method: algorithm}
    Input a $n$-qubit quantum state $\ket{\psi} \in \mathbb{C}^{2^n}$. Choose $M = \lceil 2^n - n \rceil $ and denote by $\theta_1,\theta_2, ..., \theta_M$ the angle indicating the direction of the measurement basis. Then we do the following: \\
\begin{itemize}
    \item Measure each qubit of $\ket{\psi}$ in computational basis (Z basis) $\mathcal{O}(1/\epsilon^2)$ times. 
    \item Measure the first qubit of $\ket{\psi}$ in basis $\{ \theta_i \}_{i=1}^M$ $\mathcal{O}(1/\epsilon^2)$ times. 
    \item Form the following system of nonlinear algebraic equations (e.g., as in \ref{eqn: nonlinearequation}):
    \begin{align}
    \begin{cases}
        \sum_{i_1,...,i_{n-1} = \{0,1\} } |a_{0 i_1 i_2 ... i_{n-1}}|^2  = p_0^1 \\
        \sum_{i_1,...,i_{n-1} = \{0,1\} } |a_{1 i_1 i_2 ... i_{n-1}}|^2 = p_1^1  \\
        \sum_{i_1,...,i_{n-1} = \{0,1\} } |a_{i_1 0 i_2 ..., i_{n-1}}|^2 = p_0^2 \\
        \sum_{i_1,...,i_{n-1} = \{0,1\} } |a_{i_1 1 i_2 ..., i_{n-1}}|^2 = p_1^2 \\
        \vdots \\
        \sum_{i_1,...,i_{n-1} = \{0,1\} } |a_{i_1 i_2 ..., i_{n-1} 0}|^2 = p_0^n \\
        \sum_{i_1,...,i_{n-1} = \{0,1\} } |a_{i_1 i_2 ..., i_{n-1} 1}|^2 = p_1^n \\
        \sum_{i_1,...,i_{n-1} = \{0,1\} } | \cos(\theta_1) a_{0 i_1...i_{n-1}} + \sin(\theta_1) a_{1 i_1 ... i_{n-1} }|^2 = p(\theta_1)_0 \\
        \sum_{i_1,...,i_{n-1} = \{0,1\} } | \cos(\theta_1) a_{0 i_1...i_{n-1}} - \sin(\theta_1) a_{1 i_1 ... i_{n-1} }|^2 = p(\theta_1)_1 \\
        \vdots \\
        \sum_{i_1,...,i_{n-1} = \{0,1\} } | \cos(\theta_M) a_{0 i_1...i_{n-1}} + \sin(\theta_M) a_{1 i_1 ... i_{n-1} }|^2 = p(\theta_M)_0 \\
        \sum_{i_1,...,i_{n-1} = \{0,1\} } | \cos(\theta_M)  a_{0 i_1...i_{n-1}} - \sin(\theta_M) a_{1 i_1 ... i_{n-1} }|^2 = p(\theta_M)_1 
    \end{cases}
    \label{Nnonlinearalgebraic}
    \end{align}
    \item Solve the above equations and obtain $\Re(a_{0 i_1...i_{n-1} })$, $\Im(a_{0 i_1...i_{n-1} })$, $\Re(a_{1 i_1...i_{n-1} })$, $\Im(a_{1 i_1...i_{n-1} })$ for all $i_1,i_2,...,i_{n-1} = \{0,1\}$ 
\end{itemize}
\end{method}

The last and very important point we need to discuss is the measurement-induced error and how it affects the solution we solve from such erroneous equations. Recall that the right-hand side of Eqn. \ref{Nnonlinearalgebraic} are estimated by collecting measurement results from single-qubit measurement. From an earlier discussion, we knew that by obeying the Bernoulli distribution, $\mathcal{O}( 1/\epsilon^2)$ is necessary and sufficient to estimate, for example, $p_0^1, p_1^1$ to accuracy $\epsilon$ if we measure the first qubit. Likewise, all remaining probabilities are estimated up to $\epsilon$. As the right-hand side is erroneous, the solution to the system of nonlinear algebraic equations is also deviating from the real, ideal solution. The following theorem characterizes the error deviation of the erroneous system from the ideal system: 
\begin{theorem}[Erroneous Nonlinear Algerbraic Equations]
\label{thm: erroneousnonlinearsystem}
    Let $x_1,x_2,...,x_N$ be variables and $f_1, f_2,...,f_N : \mathbb{R}^N \longrightarrow \mathbb{R}$ be multivariate functions. Define two systems of nonlinear algebraic equations as follows.
    \begin{align}
    \label{f}
        \begin{cases}
            f_1 (x_1,x_2,...,x_N) = b_1 \\
            f_2 (x_1,x_2,...,x_N) = b_2 \\
            \vdots \\
            f_N (x_1,x_2,...,x_N) = b_N 
        \end{cases}
    \end{align}
    and 
    \begin{align}
    \label{tildef}
        \begin{cases}
            f_1 (x_1,x_2,...,x_N) = \Tilde{b}_1 \\
            f_2 (x_1,x_2,...,x_N) = \Tilde{b}_2 \\
            \vdots \\
            f_N (x_1,x_2,...,x_N) = \Tilde{b}_N
        \end{cases}
    \end{align}
    Let the Jacobian be: 
    \begin{align}
        J = \begin{pmatrix}
            \frac{\partial f_1}{\partial x_1} & \frac{\partial f_1}{\partial x_2} & \cdots & \frac{\partial f_1}{\partial x_N } \\
            \frac{\partial f_2}{\partial x_1} & \frac{\partial f_2}{\partial x_2} & \cdots & \frac{\partial f_2}{\partial x_N } \\
            \cdots & \cdots & \ddots & \cdots \\
            \frac{\partial f_N}{\partial x_1} & \frac{\partial f_N}{\partial x_2} & \cdots & \frac{\partial f_N}{\partial x_N } \\
        \end{pmatrix}
    \end{align}
    Let $\textbf{x} = ( \textbf{x}_1, \textbf{x}_2,..., \textbf{x}_N)$ denote the solution to \ref{f}, and $\Tilde{\textbf{x}} = (\Tilde{\textbf{x}}_1,\Tilde{\textbf{x}}_2, ..., \Tilde{\textbf{x}}_N)$ denotes the solution to \ref{tildef}; let $\Tilde{\textbf{b}} = (\Tilde{b}_1,\Tilde{b}_2,...,\Tilde{b}_N)$ and $\textbf{b} = (b_1,b_2,...,b_N)$. Then we have that:
    \begin{align}
        | \Tilde{\textbf{x}} - \textbf{x} | \leq |J^{-1}| | \Tilde{\textbf{b}} - \textbf{b} |
    \end{align}
    where $|.|$ refers to the usual $l_2$ Euclidean norm for a vector and matrix norm for a matrix. 
\end{theorem}
\textit{Proof:} We use Taylor's expansion around the solution, in the limit $\epsilon \longrightarrow 0$. We have that, for $i=1,2,...,N$:
\begin{align}
   & f_i( \Tilde{\textbf{x}}_1,\Tilde{\textbf{x}}_2,...,\Tilde{\textbf{x}}_N) = f_i( \textbf{x}_1,\textbf{x}_2,..., \textbf{x}_N ) + \frac{\partial f_i}{\partial \textbf{x}_1} (\Tilde{\textbf{x}}_1 - \textbf{x}_1) + \frac{\partial f_i}{\partial \textbf{x}_2} (\Tilde{\textbf{x}}_2 - \textbf{x}_2) + ... + \frac{\partial f_i}{\partial \textbf{x}_N} (\Tilde{\textbf{x}}_N - \textbf{x}_N )  \\
   & \rightarrow \Tilde{b}_i = b_i + \frac{\partial f_i}{\partial \textbf{x}_1} (\Tilde{\textbf{x}}_1 - \textbf{x}_1) + \frac{\partial f_i}{\partial \textbf{x}_2} (\Tilde{\textbf{x}}_2 - \textbf{x}_2) + ... + \frac{\partial f_i}{\partial \textbf{x}_N} (\Tilde{\textbf{x}}_N - \textbf{x}_N ) 
\end{align}
Then we have that:
\begin{align}
    &\Tilde{b}_1 - b_1 = \frac{\partial f_1}{\partial \textbf{x}_1} (\Tilde{\textbf{x}}_1 - \textbf{x}_1) + \frac{\partial f_1}{\partial \textbf{x}_2} (\Tilde{\textbf{x}}_2 - \textbf{x}_2) + ... + \frac{\partial f_1}{\partial \textbf{x}_N} (\Tilde{\textbf{x}}_N - \textbf{x}_N )  \\
    &\Tilde{b}_2 - b_2 = \frac{\partial f_2}{\partial \textbf{x}_1} (\Tilde{\textbf{x}}_1 - \textbf{x}_1) + \frac{\partial f_2}{\partial \textbf{x}_2} (\Tilde{\textbf{x}}_2 - \textbf{x}_2) + ... + \frac{\partial f_2}{\partial \textbf{x}_N} (\Tilde{\textbf{x}}_N - \textbf{x}_N ) \\
    & \vdots \\
    & \Tilde{b}_N - b_N = \frac{\partial f_N}{\partial \textbf{x}_1} (\Tilde{\textbf{x}}_1 - \textbf{x}_1) + \frac{\partial f_N}{\partial \textbf{x}_2} (\Tilde{\textbf{x}}_2 - \textbf{x}_2) + ... + \frac{\partial f_N}{\partial \textbf{x}_N} (\Tilde{\textbf{x}}_N - \textbf{x}_N ) 
\end{align}
which has equivalent matrix form:
\begin{align}
    &\begin{pmatrix}
        \Tilde{b}_1 - b_1 \\
        \Tilde{b}_2 - b_2 \\
        \vdots \\
        \Tilde{b}_N - b_N 
    \end{pmatrix} = \begin{pmatrix}
            \frac{\partial f_1}{\partial x_1} & \frac{\partial f_1}{\partial x_2} & \cdots & \frac{\partial f_1}{\partial x_N } \\
            \frac{\partial f_2}{\partial x_1} & \frac{\partial f_2}{\partial x_2} & \cdots & \frac{\partial f_2}{\partial x_N } \\
            \cdots & \cdots & \ddots & \cdots \\
            \frac{\partial f_N}{\partial x_1} & \frac{\partial f_N}{\partial x_2} & \cdots & \frac{\partial f_N}{\partial x_N } \\
        \end{pmatrix} \cdot \begin{pmatrix}
            \Tilde{x}_1 - x_1 \\
        \Tilde{x}_2 - x_2 \\
        \vdots \\
        \Tilde{x}_N - x_N 
        \end{pmatrix} \\
\end{align}
which could also be written as:
\begin{align}
    \Tilde{\textbf{b}} - \textbf{b} = J ( \Tilde{\textbf{x}} - \textbf{x} )
\end{align}
Therefore, we have that:
\begin{align}
     ( \Tilde{\textbf{x}} - \textbf{x} ) = J^{-1} ( \Tilde{\textbf{b}} - \textbf{b}  )\\
     \rightarrow | ( \Tilde{\textbf{x}} - \textbf{x} ) | = | J^{-1} ( \Tilde{\textbf{b}} - \textbf{b}  ) |
     \label{eqn: 86}
\end{align}
where $|.|$ refers to the usual $l_2$ Euclidean norm. As $\Tilde{\textbf{b}} - \textbf{b} = | \Tilde{\textbf{b}} - \textbf{b}| \big( ( \Tilde{\textbf{b}} - \textbf{b})/|\Tilde{\textbf{b}} - \textbf{b}| \big)$, then we have that: 
\begin{align}
    | J^{-1} (\Tilde{\textbf{b}} - \textbf{b}) | \leq | \Tilde{\textbf{b}} - \textbf{b} | |J^{-1} |
\end{align}
As by definition, $| J^{-1}| = \max_{ y} |J^{-1} y| $ for $y$ being a unit vector. So we have that:
\begin{align}
    |  \Tilde{\textbf{x}} - \textbf{x}  | \leq |J^{-1} | | \Tilde{\textbf{b}} - \textbf{b} |
\end{align}
The proof is then completed. \\

The application of the above theorem to our main equation Eqn. \ref{Nnonlinearalgebraic} is straightforward. Such system forms a nonlinear algebraic equations, with $\Re/\Im (a_{0 i_1...i_{n-1} })$, $\Re/\Im (a_{1 i_1...i_{n-1}})$ ($i_1,...,i_{n-1} = \{0,1\} $) being variables as $(x_1,x_2,...,x_N)$ in Theorem \ref{thm: erroneousnonlinearsystem}, for which in this case we have $N = 2^{n+1}$. Additionally, the probabilities on the right hand side of equation \ref{Nnonlinearalgebraic} corresponds exactly to $\textbf{b} = (b_1,b_2,...,b_N)$ in theorem \ref{thm: erroneousnonlinearsystem}. As each probability on the right hand side of equation \ref{Nnonlinearalgebraic} is estimated up to accuracy $\epsilon$, it translates directly to the fact that for all $i=1,2,...,N$, $|\Tilde{b}_i - b_i| \leq \epsilon$, then we have that:
\begin{align}
    | \Tilde{\textbf{b}} - \textbf{b} | = \sqrt{ \sum_{i=1}^N (\Tilde{b}_i - b_i)^2 } \leq  \sqrt{N} \epsilon
\end{align}
Therefore, from the above, we have:
\begin{align}
    | \Tilde{\textbf{x}} - \textbf{x} | \leq |J^{-1} |  \sqrt{N}\epsilon
    \label{90}
\end{align}
where we remind that, $\textbf{x}$ is being used in an abusive way (following theorem \ref{thm: erroneousnonlinearsystem}) to denote the ideal solution to our main system of interest \ref{Nnonlinearalgebraic}, and $\Tilde{\textbf{x}}$ denotes the erroneous solution. The above equation suggests that the error deviation depends on the behavior of corresponding Jacobian $J$. If $J$ has the lowest non-zero singular value being $\mathcal{O}(1/poly(N))$ then the norm $|J^{-1}|$ gonna be large, making error induced larger. Otherwise, if the lowest non-zero singular value being large enough, then $|J^{-1}|$ could be $\mathcal{O}(1)$.  Then $| \Tilde{\textbf{x}}-\textbf{x} | \leq \mathcal{O}( 2 \sqrt{N} \epsilon)$. Now we discuss a few specific scenarios to show how our framework can be applied in a broader context. 

\subsubsection{Estimating each amplitude and its norm with additive error $\delta$ }
\label{sec: additiveerror}
Suppose our $n$-qubit state is $\ket{\psi} = \sum_{i_0,i_1,...,i_{n-1} = {0,1} } a_{i_0i_1...i_{n-1}} \ket{i_0 i_1... i_{n-1}}$ where each $a_{i_0i_1...i_{n-1}}$ is a complex number, and we are interested in estimating its norm $| a_{i_0 i_1 ...i_{n-1}}|$ to additive error $\epsilon$, for all amplitudes. As it is a complex number, we have that:
\begin{align}
    |a_{i_0 i_1 ... i_{n-1}}| = \sqrt{\Re(a_{i_0 i_1 ... i_{n-1}})^2 + \Im( a_{i_0 i_1 ... i_{n-1}})^2 }
\end{align}
The framework we introduced return an approximation to real and imaginary component of all amplitudes $a_{i_0i_1...i_{n-1}}$. As we expect all norm of amplitudes having additive accuracy $\delta$, it is equivalent to:
\begin{align}
    \max_{i_0,i_1,...,i_{n-1}}  \{ \big| |\Tilde{a}_{i_0i_1...i_{n-1}}| - |a_{i_0i_1...i_{n-1}}|\big|  \} = \delta
\end{align}
where $\Tilde{a}_{i_0i_1...i_{n-1}}$ refers to the approximation of corresponding amplitude $a_{i_0i_1...i_{n-1}}$. Let $\Re(\Tilde{a}_{i_0i_1...i_{n-1}})$ and $\Im(\Tilde{a}_{i_0i_1...i_{n-1}})$ denotes the approximation of real and imaginary part of $a_{i_0i_1...i_{n-1}}$. Let $\Tilde{a}_{i_0i_1...i_{n-1}} = \Re(\Tilde{a}_{i_0i_1...i_{n-1}}) + i\Im(\Tilde{a}_{i_0i_1...i_{n-1}} ) $. We have that:
\begin{align}
    \big| |\Tilde{a}_{i_0i_1...i_{n-1}}| - |a_{i_0i_1...i_{n-1}}| \big|^2 &= \big| \sqrt{\Re(\Tilde{a}_{i_0i_1...i_{n-1}})^2 +\Im(\Tilde{a}_{i_0i_1...i_{n-1}})^2  } - \sqrt{\Re(a_{i_0 i_1 ... i_{n-1}})^2 + \Im( a_{i_0 i_1 ... i_{n-1}})^2 } \big|^2 \\
    &\leq \big| \Re(\Tilde{a}_{i_0i_1...i_{n-1}})^2 +\Im(\Tilde{a}_{i_0i_1...i_{n-1}})^2  - ( \Re(a_{i_0 i_1 ... i_{n-1}})^2 + \Im( a_{i_0 i_1 ... i_{n-1}})^2 )  \big| \\
    &\leq \big| \big( \Re(\Tilde{a}_{i_0i_1...i_{n-1}})^2 -  \Re(a_{i_0 i_1 ... i_{n-1}})^2 \big) + \big( \Im(\Tilde{a}_{i_0i_1...i_{n-1}})^2 -\Im( a_{i_0 i_1 ... i_{n-1}})^2   \big)   \big| \\
    &\leq \big|  | \Re(\Tilde{a}_{i_0i_1...i_{n-1}})^2 -  \Re(a_{i_0 i_1 ... i_{n-1}})^2 | + | \Im(\Tilde{a}_{i_0i_1...i_{n-1}})^2 -\Im( a_{i_0 i_1 ... i_{n-1}})^2   |   \big|
\end{align}
where the second line comes from the fact that for any $x,y$, we have $(x-y)^2 = (x-y)(x+y) \rightarrow (x-y)^2 = |(x-y)^2 | = | (x-y) (x-y) | \leq | (x-y)(x+y) | \leq |x^2 - y^2|$. Remind that we are seeking for:
\begin{align}
    \max_{i_0,i_1,...,i_{n-1}} \big| \big( \Re(\Tilde{a}_{i_0i_1...i_{n-1}})^2 -  \Re(a_{i_0 i_1 ... i_{n-1}})^2 \big) + \big( \Im(\Tilde{a}_{i_0i_1...i_{n-1}})^2 -\Im( a_{i_0 i_1 ... i_{n-1}})^2   \big)   \big| \leq \epsilon^2 
\end{align}
and remind further that all the values $\Re(\Tilde{a}_{i_0i_1...i_{n-1}}), \Im(\Tilde{a}_{i_0i_1...i_{n-1}})$, and $\Re(a_{i_0 i_1 ... i_{n-1}}), \Im( a_{i_0 i_1 ... i_{n-1}})$ (for all $i_0,i_1,...i_{n-1}$) are contained inside $\Tilde{\textbf{x}}$ and $\textbf{x}$, as we denoted. We have that:
\begin{align}
   |  \Re(\Tilde{a}_{i_0i_1...i_{n-1}})^2-  \Re(a_{i_0 i_1 ... i_{n-1}})^2 | &= |  (\Re(\Tilde{a}_{i_0i_1...i_{n-1}}) - \Re(a_{i_0 i_1 ... i_{n-1}})   )(\Tilde{\Re}(a_{i_0i_1...i_{n-1}}) + \Re(a_{i_0 i_1 ... i_{n-1}}) ) | \\
   &\leq 2 |\Re(\Tilde{a}_{i_0i_1...i_{n-1}}) - \Re(a_{i_0 i_1 ... i_{n-1}})   |
\end{align}
where we have used the fact that $|\Re(\Tilde{a}_{i_0i_1...i_{n-1}})|, |\Re(a_{i_0 i_1 ... i_{n-1}})| \leq 1 $. Likewise:
\begin{align}
    | \Im(\Tilde{a}_{i_0i_1...i_{n-1}})^2 -\Im( a_{i_0 i_1 ... i_{n-1}})^2   | \leq 2  |  \Im(\Tilde{a}_{i_0i_1...i_{n-1}}) - \Im( a_{i_0 i_1 ... i_{n-1}})  | 
\end{align}
So we have:
\begin{align}
    \big|  | \Re(\Tilde{a}_{i_0i_1...i_{n-1}})^2 -  \Re(a_{i_0 i_1 ... i_{n-1}})^2 | + | \Im(\Tilde{a}_{i_0i_1...i_{n-1}})^2 -\Im( a_{i_0 i_1 ... i_{n-1}})^2   |   \big| \leq & 2 | \Re(\Tilde{a}_{i_0i_1...i_{n-1}}) - \Re(a_{i_0 i_1 ... i_{n-1}})   | + \\ &2  |  \Im(\Tilde{a}_{i_0i_1...i_{n-1}})-\Im( a_{i_0 i_1 ... i_{n-1}})  | 
\end{align}
We have the Cauchy-Schwarz inequality:
\begin{align}
 & | \Re(\Tilde{a}_{i_0i_1...i_{n-1}}) - \Re(a_{i_0 i_1 ... i_{n-1}})   | + |  \Im(\Tilde{a}_{i_0i_1...i_{n-1}}) -\Im( a_{i_0 i_1 ... i_{n-1}})  |  \\ & \leq \sqrt{2 \big(  | \Re(\Tilde{a}_{i_0i_1...i_{n-1}}) - \Re(a_{i_0 i_1 ... i_{n-1}})   |^2 + |  \Im(\Tilde{a}_{i_0i_1...i_{n-1}}) - \Im( a_{i_0 i_1 ... i_{n-1}})  |^2   \big) }
\end{align}
Therefore, for all $i_0,i_1,...,i_{n-1}$, we have:
\begin{align}
    & \big| \big( \Re(\Tilde{a}_{i_0i_1...i_{n-1}})^2 -  \Re(a_{i_0 i_1 ... i_{n-1}})^2 \big) + \big( \Im(\Tilde{a}_{i_0i_1...i_{n-1}})^2 -\Im( a_{i_0 i_1 ... i_{n-1}})^2   \big)   \big| 
    \\ &\leq 2\sqrt{2 \big(  | \Re(\Tilde{a}_{i_0i_1...i_{n-1}}) - \Re(a_{i_0 i_1 ... i_{n-1}})   |^2 + |  \Im(\Tilde{a}_{i_0i_1...i_{n-1}}) - \Im( a_{i_0 i_1 ... i_{n-1}})  |^2   \big) }
\end{align}
Recall that from previous discussion that lead to equation \ref{90}, we have that:
\begin{align}
    | \Tilde{\textbf{x}} - \textbf{x} | \leq |J^{-1} | \sqrt{N}\epsilon
\end{align}
where $\textbf{x}$ contains the ideal solution $\Re(a_{i_0 i_1 ... i_{n-1}}), \Im( a_{i_0 i_1 ... i_{n-1}}) $ and $\Tilde{\textbf{x}}$ contains the approximated ones, for all $i_0,i_1,...,i_{n-1} = \{0,1\}$. Hence, $\Tilde{\textbf{x}} - \textbf{x}$ is a vector containing all $(  \Re(\Tilde{a}_{i_0i_1...i_{n-1}}) - \Re(a_{i_0 i_1 ... i_{n-1}})  )$ and $\Im(\Tilde{a}_{i_0i_1...i_{n-1}})-\Im( a_{i_0 i_1 ... i_{n-1}})  $. It is apparent that, for any specific $i_0i_1...i_{n-1}$, we have:
\begin{align}
    & \big(  | \Re(\Tilde{a}_{i_0i_1...i_{n-1}}) - \Re(a_{i_0 i_1 ... i_{n-1}})   |^2 + |  \Im(\Tilde{a}_{i_0i_1...i_{n-1}}) - \Im( a_{i_0 i_1 ... i_{n-1}})  |^2   \big)  \\ & \leq \sum_{i_0,i_1,...,i_{n-1}} \big(  | \Re(\Tilde{a}_{i_0i_1...i_{n-1}}) - \Re(a_{i_0 i_1 ... i_{n-1}})   |^2 + |  \Im(\Tilde{a}_{i_0i_1...i_{n-1}})-\Im( a_{i_0 i_1 ... i_{n-1}})  |^2   \big)
    \\ & = |\Tilde{\textbf{x}} - \textbf{x}|^2 \leq  |J^{-1}|^2 N \epsilon^2 
\end{align}
Therefore:
\begin{align}
    & \sqrt{2 \big(  | \Re(\Tilde{a}_{i_0i_1...i_{n-1}})^2 - \Re(a_{i_0 i_1 ... i_{n-1}})   |^2 + |  \Im(\Tilde{a}_{i_0i_1...i_{n-1}}) - \Im( a_{i_0 i_1 ... i_{n-1}})  |^2   \big) } 
    \\&\leq \sqrt{2}  |J^{-1}| \sqrt{N} \epsilon  \\
\end{align}
It implies that:
\begin{align}
    & \max_{i_0,i_1,...,i_{n-1}} \big| \Re(\Tilde{a}_{i_0i_1...i_{n-1}})^2 -  \Re(a_{i_0 i_1 ... i_{n-1}})^2 \big) + \big( \Im(\Tilde{a}_{i_0i_1...i_{n-1}})^2 -\Im( a_{i_0 i_1 ... i_{n-1}})^2   \big)   \big|  
    \\ &\leq 2\sqrt{2 \big(  | \Re(\Tilde{a}_{i_0i_1...i_{n-1}}) - \Re(a_{i_0 i_1 ... i_{n-1}})   |^2 + |  \Im(\Tilde{a}_{i_0i_1...i_{n-1}}) - \Im( a_{i_0 i_1 ... i_{n-1}})  |^2   \big) } 
    \\ &\leq 2 \sqrt{2} |J^{-1}| \sqrt{N}\epsilon
\end{align}
Previously, we also showed that:
\begin{align}
    \big| |\Tilde{a}_{i_0i_1...i_{n-1}}| - |a_{i_0i_1...i_{n-1}}| \big|^2 &\leq \big|  | \Re(\Tilde{a}_{i_0i_1...i_{n-1}})^2 -  \Re(a_{i_0 i_1 ... i_{n-1}})^2 | + | \Im(\Tilde{a}_{i_0i_1...i_{n-1}})^2 -\Im( a_{i_0 i_1 ... i_{n-1}})^2   |   \big| \\ 
    \longrightarrow  \big| |\Tilde{a}_{i_0i_1...i_{n-1}}| - |a_{i_0i_1...i_{n-1}}| \big| &\leq \sqrt{\big|  | \Re(\Tilde{a}_{i_0i_1...i_{n-1}})^2 -  \Re(a_{i_0 i_1 ... i_{n-1}})^2 | + | \Im(\Tilde{a}_{i_0i_1...i_{n-1}})^2 -\Im( a_{i_0 i_1 ... i_{n-1}})^2   |   \big| }
\end{align}
Hence:
\begin{align}
    \max_{i_0,i_1,...,i_{n-1}} \big| |\Tilde{a}_{i_0i_1...i_{n-1}}| - |a_{i_0i_1...i_{n-1}}| \big| &\leq \max_{i_0,i_1,...,i_{n-1}} \sqrt{\big|  |\Re(\Tilde{a}_{i_0i_1...i_{n-1}})^2 -  \Re(a_{i_0 i_1 ... i_{n-1}})^2 | + | \Im(\Tilde{a}_{i_0i_1...i_{n-1}})^2 -\Im( a_{i_0 i_1 ... i_{n-1}})^2   |   \big|} \\
    &\leq  8^{1/4} \sqrt{J^{-1}} \sqrt{\sqrt{N}} \sqrt{\epsilon}
\end{align}
We expect $\max_{i_0,i_1,...,i_{n-1}} \big| |\Tilde{a}_{i_0i_1...i_{n-1}}| - |a_{i_0i_1...i_{n-1}}| \big| \leq \delta $ so it is sufficient to set:
\begin{align}
    2 2^{1/4} \sqrt{J^{-1}} \sqrt{\sqrt{N}} \sqrt{\epsilon} = \delta \\
    \longrightarrow \epsilon = \frac{\delta^2}{ 4\sqrt{2} |J^{-1}| \sqrt{N}}
\end{align}
Per algorithm \ref{method: algorithm}, we first make measurement on each qubit $\mathcal{O}(1/\epsilon^2)$ times, following by measuring first qubit $\mathcal{O}( 2^n-n)$ times. Given that $N = 2^{n+1}$, the total number of measurement (or equivalently, the number of sample of $n$-qubit state $\ket
\psi$) required to estimate all amplitudes to additive error $\delta$ is  $ \mathcal{O}\big( \frac{4^n }{\delta^4}   \big)  $. 

The previous discussion was devoted to estimating the norm with maximum precision $\delta$. Now we discuss a quite similar goal, which is outputting amplitudes such that:
\begin{align}
    \max_{i_0,i_1,...,i_{n-1}}  \{ \big| \Tilde{a}_{i_0i_1...i_{n-1}}- a_{i_0i_1...i_{n-1}}|\big|  \} = \delta
\end{align}
We have that:
\begin{align}
    \big| \Tilde{a}_{i_0i_1...i_{n-1}}- a_{i_0i_1...i_{n-1}}|\big| &= \sqrt{ ( \Re(\Tilde{a}_{i_0i_1...i_{n-1}}) - \Re( a_{i_0i_1...i_{n-1}} )  )^2 + (  \Im(\Tilde{a}_{i_0i_1...i_{n-1}}) - \Im( a_{i_0 i_1 ... i_{n-1}})  )^2    }  \\
    & \leq \sqrt{ \sum_{i_0,i_1,...,i_{n-1}} \big(  | \Re(\Tilde{a}_{i_0i_1...i_{n-1}}) - \Re(a_{i_0 i_1 ... i_{n-1}})   |^2 + |  \Im(\Tilde{a}_{i_0i_1...i_{n-1}})-\Im( a_{i_0 i_1 ... i_{n-1}})  |^2   \big)} \\
    &= |\Tilde{ \textbf{x} } - \textbf{x}| \leq |J^{-1}| \sqrt{2^n}\epsilon = \delta
\end{align}
which directly implies that $\epsilon = \delta/ (|J^{-1}| \sqrt{2^n} )$. So, the total number of measurements is $\mathcal{O}(2^n/\epsilon^2) = \mathcal{O}( |J^{-1}| 4^n/\delta^4)$.  \\

Naively, as we have discussed, the common approach makes use of jointly all qubit measurement to estimate the amplitude square. More concretely, let $\ket{\psi} = \sum_{i_0,i_1,...,i_{n-1} = \{0,1\} } a_{i_0i_1...i_{n-1}} \ket{i_0 i_1... i_n-1}$. Suppose that we want to estimate $a_{i_0i_1...i_{n-1}}$ for a specific $i_0,i_1,...,i_{n-1}$. Then we perform the measurement on all qubits, and the outcome forms a Bernoulli distribution for $i_0 i_1 ... i_{n-1}$ and the remaining strings, with corresponding probability $|a_{i_0i_1...i_{n-1}}|^2$ and $1- |a_{i_0i_1...i_{n-1}}|^2$. Then it takes $\mathcal{O}(1/\epsilon^2)$ measurements to estimate this probability $ |a_{i_0i_1...i_{n-1}}|^2$ to error $\epsilon$. The procedure is then repeated to estimate all amplitude squares $|a_{i_0i_1...i_{n-1}}|^2$ to error $\epsilon$, resulting in $\mathcal{O}(2^n/\epsilon^2)$ measurements in total. However, we note that these estimations are for the amplitude square. If we want to find the amplitude, we need to take the square root, with further error propagation as follows. Let $|\Tilde{a}_{i_0i_1...i_{n-1}}|^2$ denotes the amplitude denote the approximation to amplitude square of $|a_{i_0i_1...i_{n-1}}|^2$, i.e:
\begin{align}
    \big| |\Tilde{a}_{i_0i_1...i_{n-1}}|^2 -|a_{i_0i_1...i_{n-1}}|^2 \big| \leq \epsilon
\end{align}
Then we have that:
\begin{align}
    \big| \sqrt{|\Tilde{a}_{i_0i_1...i_{n-1}}|^2 } - \sqrt{|a_{i_0i_1...i_{n-1}}|^2  }   \big|^2 & \leq \big| \sqrt{|\Tilde{a}_{i_0i_1...i_{n-1}}|^2 } - \sqrt{|a_{i_0i_1...i_{n-1}}|^2  }   \big| \big| \sqrt{|\Tilde{a}_{i_0i_1...i_{n-1}}|^2 } + \sqrt{|a_{i_0i_1...i_{n-1}}|^2  }   \big| \\
    & = \big| |\Tilde{a}_{i_0i_1...i_{n-1}}|^2 -|a_{i_0i_1...i_{n-1}}|^2 \big| \leq \epsilon
\end{align}
Therefore, if we expect the error to be $\delta$, we need to set $\epsilon = \delta^2$:
\begin{align}
    \big| \sqrt{|\Tilde{a}_{i_0i_1...i_{n-1}}|^2 } - \sqrt{|a_{i_0i_1...i_{n-1}}|^2  }   \big|^2 \leq \delta^2 \\
    \rightarrow \big| \sqrt{|\Tilde{a}_{i_0i_1...i_{n-1}}|^2 } - \sqrt{|a_{i_0i_1...i_{n-1}}|^2  }   \big| \leq \delta 
\end{align}
and the total number of measurement needed is $\mathcal{O}(2^n/\delta^4)$, which is quadratically better with respect to $2^n$ comparing to $\mathcal{O}(4^n/\delta^4) $ measurements required by the method introduced, which is quite expected because naive procedure employs more qubits measurement.

\subsubsection{Estimating probability distribution with total variation $\delta$}
\label{sec: totalvariance}
Suppose our $n$-qubit state is $\ket{\psi} = \sum_{i_0,i_1,...,i_{n-1} = {0,1} } a_{i_0i_1...i_{n-1}} \ket{i_0 i_1... i_{n-1}}$, if we measure all qubits in computational basis, then apparently the outcome $i$ with corresponding probability $p(i_0i_1...i_{n-1}) = |a_{i_0i_1...i_{n-1}}|^2$ forms a probability distribution, denoted as $p$. If we wish to estimate such a distribution up to total variation $\delta$, that is, obtain a distribution $\Tilde{p}$ such that $|\Tilde{p} - p|_1 = \sum_{i_0,i_1,...i_{n-1} = 0,1} |\Tilde{p}(i_0i_1...i_{n-1}) - p(i_0i_1...i_{n-1}) | \leq \delta$. Similarly to the previous discussion, let $\Re(a_i)$ and $\Im(a_i)$ denote the real and imaginary components of $a_i$, and $\Re(\Tilde{a}_i)/ \Im(\Tilde{a}_i)$ denotes the approximation to the corresponding real and imaginary components. We want that:
\begin{align}
    \sum_{i_0,i_1,...,i_{n-1} = {0,1} } | \Re(\Tilde{a}_{i_0i_1...i_{n-1}})^2 + \Im(\Tilde{a}_{i_0i_1...i_{n-1}})^2 - \Re(a_{i_0i_1...i_{n-1}})^2 - \Im(a_{i_0i_1...i_{n-1}})^2 | &  \leq \delta
\end{align}
As discussed above, we have that:
\begin{align}
    & | \Re(\Tilde{a}_{i_0i_1...i_{n-1}})^2 + \Im(\Tilde{a}_{i_0i_1...i_{n-1}})^2 - \Re(a_{i_0i_1...i_{n-1}})^2 - \Im(a_{i_0i_1...i_{n-1}})^2 | \\ &\leq | \Re(\Tilde{a}_{i_0i_1...i_{n-1}})^2- \Re(a_{i_0i_1...i_{n-1}})^2| +  |\Im(\Tilde{a}_{i_0i_1...i_{n-1}})^2- \Im(a_{i_0i_1...i_{n-1}})^2 | \\
    & \leq 2 | \Re(\Tilde{a}_{i_0i_1...i_{n-1}})  - \Re(a_{i_0i_1...i_{n-1}})| + 2 | \Im(\Tilde{a}_{i_0i_1...i_{n-1}})  -   \Im(a_{i_0i_1...i_{n-1}})| \\ 
\end{align}
Therefore:
\begin{align}
    & \sum_{i_0,i_1,...,i_{n-1} = {0,1} } | \Re(\Tilde{a}_{i_0i_1...i_{n-1}})^2 + \Im(\Tilde{a}_{i_0i_1...i_{n-1}})^2 - \Re(a_{i_0i_1...i_{n-1}})^2 - \Im(a_{i_0i_1...i_{n-1}})^2 | \\ & \leq 2 \sum_{i_0,i_1,...,i_{n-1} = {0,1} } \big(  ( \Re(\Tilde{a}_{i_0i_1...i_{n-1}})  - \Re(a_{i_0i_1...i_{n-1}})) + ( \Im(\Tilde{a}_{i_0i_1...i_{n-1}})  -   \Im(a_{i_0i_1...i_{n-1}}) \big) \\
    \\ & \leq 2   \sqrt{ \sum_{i_0,i_1,...,i_{n-1} = {0,1} } 2^{n+1} \big( ( \Re(\Tilde{a}_{i_0i_1...i_{n-1}})  - \Re(a_{i_0i_1...i_{n-1}}))^2 + ( \Im(\Tilde{a}_{i_0i_1...i_{n-1}})  -   \Im(a_{i_0i_1...i_{n-1}}) )^2    \big) }
\end{align}
where the second line comes from Cauchy-Schwarz inequality. Recall that we have:
\begin{align}
    | \Tilde{\textbf{x}} - \textbf{x} | &=   \sqrt{ \sum_{i_0,i_1,...,i_{n-1} = {0,1} }  \big( ( \Re(\Tilde{a}_{i_0i_1...i_{n-1}})  - \Re(a_{i_0i_1...i_{n-1}}))^2 + ( \Im(\Tilde{a}_{i_0i_1...i_{n-1}})  - \Im(a_{i_0i_1...i_{n-1}}) )^2    \big)} \\& \leq |J^{-1} |  \sqrt{N}\epsilon = |J^{-1}| \sqrt{2^n} \epsilon
\end{align}
So we have that:
\begin{align}
    \sum_{i_0,i_1,...,i_{n-1} = {0,1} } | \Re(\Tilde{a}_{i_0i_1...i_{n-1}})^2 + \Im(\Tilde{a}_{i_0i_1...i_{n-1}})^2 - \Re(a_{i_0i_1...i_{n-1}})^2 - \Im(a_{i_0i_1...i_{n-1}})^2 | &  \leq 2 \sqrt{2^{n+1}} |J^{-1}| \sqrt{2^n} \epsilon = 2\sqrt{2} 2^n |J^{-1}| \epsilon
\end{align}
As desired, we expect the total variation to be $\delta$, so we need to set $2\sqrt{2} 2^n |J^{-1}| \epsilon = \delta \rightarrow \epsilon = \delta/( 2\sqrt{2} 2^n |J^{-1}|   ) $. The total number of measurements is then $\mathcal{O}(2^n/\epsilon^2) = \mathcal{O}( 6^n |J^{-1}|/\delta^4 )$.  

In a naive approach, as described earlier, it takes in total of $\mathcal{O}(2^n/\epsilon^2)$ measurement to estimate all amplitudes square $|a_{i_0i_1...i_{n-1}}|^2$, each to accuracy $\epsilon$. Hence, the accumulated error is:
\begin{align}
    \sum_{i_0,i_1,...,i_{n-1} = {0,1} } ||\Tilde{a}_{i_0i_1...i_{n-1}}|^2 -  |a_{i_0i_1...i_{n-1}}|^2 | \leq 2^n \epsilon 
\end{align}
If we want the total variation to be $\delta$, then we require $ 2^n \epsilon = \delta \rightarrow \epsilon = \delta/2^n$. So the total measurement required is $\mathcal{O}(4^n/\epsilon^2)$. We remark that, given the decomposition of initial state $\ket{\psi}$, its density matrix representation is:
\begin{align}
    \rho = \ket{\psi}\bra{\psi} = \sum_{i_0,i_1,...,i_{n-1} = {0,1} } a_{i_0i_1...i_{n-1}} \ket{i_0 i_1... i_{n-1}} \sum_{i_0,i_1,...,i_{n-1} = {0,1} } a^*_{i_0i_1...i_{n-1}} \bra{i_0 i_1... i_{n-1}}
\end{align}
And the fact that we expect to obtain approximations of all amplitudes with total variation
\begin{align}
    \sum_{i_0,i_1,...,i_{n-1} = {0,1} } ||\Tilde{a}_{i_0i_1...i_{n-1}}|^2 -  |a_{i_0i_1...i_{n-1}}|^2 | \leq \delta
\end{align}
is essentially equivalent to obtaining a state $\sigma$ such that $| \sigma- \rho|_1 = \Tr | \sigma-\rho| \leq \delta$, e.g., defining 
$$\sigma = \sum_{i_0,i_1,...,i_{n-1} = {0,1} } |\Tilde{a}_{i_0i_1...i_{n-1}}|^2 \ket{i_0 i_1... i_{n-1}} \bra{i_0 i_1... i_{n-1}} $$
This problem has been extensively studied in \cite{yu2020sample, haah2016sample}, where the authors proposed different measurement schemes to output the desired density matrix that approximates the given state. The work \cite{haah2016sample} employs more qubits measurement, meanwhile \cite{yu2020sample} employs single-qubit measurement. The complexity of \cite{yu2020sample} is $\mathcal{O}(10^n/\delta^2)$ measurement required to output the density state to trace distance $\delta$, which is quadratically more efficient than our method in error $\delta$, but less efficient by a factor of $4^n$.

\subsubsection{Estimating with average $L_1$-norm accuracy }
\label{sec: estimatingl1norm}
Again, let $\ket{\psi} = \sum_{i_0,i_1,...,i_{n-1} = {0,1} } a_{i_0i_1...i_{n-1}} \ket{i_0 i_1... i_{n-1}}$ and instead of individual error $\delta$ as in section \ref{sec: additiveerror}, we expect the average error to be $\delta$:
\begin{align}
    \sum_{i_0,i_1,...,i_{n-1}} \frac{1}{2^n} ||\Tilde{a}_{i_0i_1...i_{n-1}}| -  |a_{i_0i_1...i_{n-1}}|| \leq \delta
\end{align}
Similar as before, let $a_{i_0i_1...i_{n-1}} = \Re(a_{i_0i_1...i_{n-1}}) + i \Im( a_{i_0i_1...i_{n-1}}) $, and $\Tilde{a}_{i_0i_1...i_{n-1}} = \Re(  \Tilde{a}_{i_0i_1...i_{n-1}}) + i \Im( \Tilde{a}_{i_0i_1...i_{n-1}} ) $ where $ \Re(  \Tilde{a}_{i_0i_1...i_{n-1}})$ and $  \Im( \Tilde{a}_{i_0i_1...i_{n-1}} )$ denotes the approximation to $\Re(a_{i_0i_1...i_{n-1}}),\Im( a_{i_0i_1...i_{n-1}})$ respectively. As showed previously, we have that:
\begin{align}
    \big| |\Tilde{a}_{i_0i_1...i_{n-1}}| - |a_{i_0i_1...i_{n-1}}| \big|^2 &\leq \big| |\Tilde{a}_{i_0i_1...i_{n-1}}|^2 -  |a_{i_0i_1...i_{n-1}}|^2  \big| \\
    &= \big| \Re(  \Tilde{a}_{i_0i_1...i_{n-1}})^2 + \Im( \Tilde{a}_{i_0i_1...i_{n-1}} )^2 -\Re(a_{i_0i_1...i_{n-1}})^2 -\Im( a_{i_0i_1...i_{n-1}})^2   \big| \\
    & \leq 2 | \Re(\Tilde{a}_{i_0i_1...i_{n-1}})  - \Re(a_{i_0i_1...i_{n-1}}) | + 2 | \Im(\Tilde{a}_{i_0i_1...i_{n-1}})  -   \Im(a_{i_0i_1...i_{n-1}}) | \\ 
    & \leq 2 \sqrt{2}\sqrt{( \Re(\Tilde{a}_{i_0i_1...i_{n-1}})  - \Re(a_{i_0i_1...i_{n-1}}))^2 + ( \Im(\Tilde{a}_{i_0i_1...i_{n-1}})  -   \Im(a_{i_0i_1...i_{n-1}}))^2 } 
\end{align}
We have:
\begin{align}
    & \frac{1}{2^n}  \sum_{i_0,i_1,...,i_{n-1}} ||\Tilde{a}_{i_0i_1...i_{n-1}}| -  |a_{i_0i_1...i_{n-1}}|| \leq \frac{1}{2^n} \sqrt{ 2^n \sum_{i_0,i_1,...,i_{n-1}  } \big| |\Tilde{a}_{i_0i_1...i_{n-1}}| - |a_{i_0i_1...i_{n-1}}| \big|^2 } \\
    & \leq \frac{1}{2^n} \sqrt{ 2^n  \sum_{i_0,i_1,...,i_{n-1}  } \sqrt{2}\sqrt{( \Re(\Tilde{a}_{i_0i_1...i_{n-1}})  - \Re(a_{i_0i_1...i_{n-1}}))^2 + ( \Im(\Tilde{a}_{i_0i_1...i_{n-1}})  -   \Im(a_{i_0i_1...i_{n-1}}))^2 }  }  \\
    & \leq \frac{1}{2^n} \sqrt{2^n \sqrt{2}  \sqrt{2^n \sum_{i_0,i_1,...,i_{n-1}  } (  ( \Re(\Tilde{a}_{i_0i_1...i_{n-1}})  - \Re(a_{i_0i_1...i_{n-1}}))^2 + ( \Im(\Tilde{a}_{i_0i_1...i_{n-1}})  -   \Im(a_{i_0i_1...i_{n-1}}))^2  )  }   } \\
    & \leq \frac{1}{2^n} \sqrt{  2^n \sqrt{2} \sqrt{2^n  |\Tilde{\textbf{x}} - \textbf{x}|^2  }  } \\
    &\leq \frac{1}{2^n} \sqrt{2^n \sqrt{2} \sqrt{2^n} |J^{-1}| \sqrt{2^n} \epsilon } = 2^{1/4} \sqrt{|J^{-1}| \epsilon} 
\end{align}
where the last line comes from Eqn. \ref{90}, with $N=2^n$. We expect that 
\begin{align}
    \sum_{i_0,i_1,...,i_{n-1}} \frac{1}{2^n} ||\Tilde{a}_{i_0i_1...i_{n-1}}| -  |a_{i_0i_1...i_{n-1}}|| \leq \delta
\end{align}
so we need to set $2^{1/4} \sqrt{|J^{-1}| \epsilon}  = \delta \rightarrow \epsilon = \delta^2/(\sqrt{2} |J^{-1}| )$. Hence, the total number of measurements needed is $\mathcal{O}(2^n/\epsilon^2) = \mathcal{O}( 2^n |J^{-1}|^2 /\delta^4 )$. Naively, we have seen that it takes $\mathcal{O}(2^n/\epsilon^2)$ number of measurements to estimate all amplitude square $|a_{i_0i_1... i_{n-1}}|^2$ to accuracy $\epsilon$. We also had from previous discussion, in section \ref{sec: additiveerror} that 
\begin{align}
    \big| \sqrt{|\Tilde{a}_{i_0i_1...i_{n-1}}|^2 } - \sqrt{|a_{i_0i_1...i_{n-1}}|^2  }   \big|^2 & \leq \big| \sqrt{|\Tilde{a}_{i_0i_1...i_{n-1}}|^2 } - \sqrt{|a_{i_0i_1...i_{n-1}}|^2  }   \big| \big| \sqrt{|\Tilde{a}_{i_0i_1...i_{n-1}}|^2 } + \sqrt{|a_{i_0i_1...i_{n-1}}|^2  }   \big| \\
    & = \big| |\Tilde{a}_{i_0i_1...i_{n-1}}|^2 -|a_{i_0i_1...i_{n-1}}|^2 \big| \leq \epsilon
\end{align}
Hence:
\begin{align}
     \big||\Tilde{a}_{i_0i_1...i_{n-1}}|  - |a_{i_0i_1...i_{n-1}}|   \big| \leq \sqrt{\epsilon} \\
     \longrightarrow \sum_{i_0,i_1,...,i_{n-1}} \frac{1}{2^n} ||\Tilde{a}_{i_0i_1...i_{n-1}}| -  |a_{i_0i_1...i_{n-1}}|| \leq \sqrt{\epsilon}
\end{align}
If we expect the average $L_1$ norm error to be $\delta$, then we need to set $\sqrt{\epsilon} = \delta \rightarrow \epsilon = \delta^2$. So using naively, jointly multi-qubits measurement approach, the total number of measurement required is $\mathcal{O}( 2^n/\epsilon^2) = \mathcal{O}(2^n /\delta^4)$. Hence, in comparison, our single-qubit measurement framework achieves assymtotically similar number of measurements, given that the norm $|J^{-1}|$ grows as $\mathcal{O}(1)$.

\section{Conclusion}
\label{sec: conclusion}
In this work, we have outlined a simple framework for estimating amplitudes of a given state, using single-qubit measurements only. Our method executes measurement on each qubit in multiple basis, then uses the outcomes to construct a system of nonlinear algebraic equations. By classically solving such equations, we are able to obtain the approximation to the desired amplitudes. Our work adds to many existing literature regarding quantum state tomography, a simple way to recover a quantum state based on single-qubit measurement. One may wonder if our protocol can perform better if, instead of single-qubit measurement, we use multiqubit measurements instead. We recall that the key point of our method is to construct a nonlinear algebraic system. Performing more multiqubit measurements, in principle, would just produce a different set of nonlinear algebraic equations. Therefore, in some sense, our protocol implies that multiple-qubit measurement is not asymptotically stronger than single-qubit measurement. The trickiest aspect of our protocol is the norm of the inverse of Jacobian of the underlying nonlinear algebraic system. We expect such a norm to grow as much as $\mathcal{O}(1)$, however, we are not able to prove it, and we believe that in general the behavior of such Jacobian can only be revealed numerically.

\bibliography{ref.bib}{}
\bibliographystyle{unsrt}

\clearpage
\newpage
\onecolumngrid
\appendix

\end{document}